# Advanced Integration Strategies for ESD Protection and Termination in High-Speed LVDS Systems: A Comprehensive Design Approach

Kavya Gaddipati

Arizona State University, USA

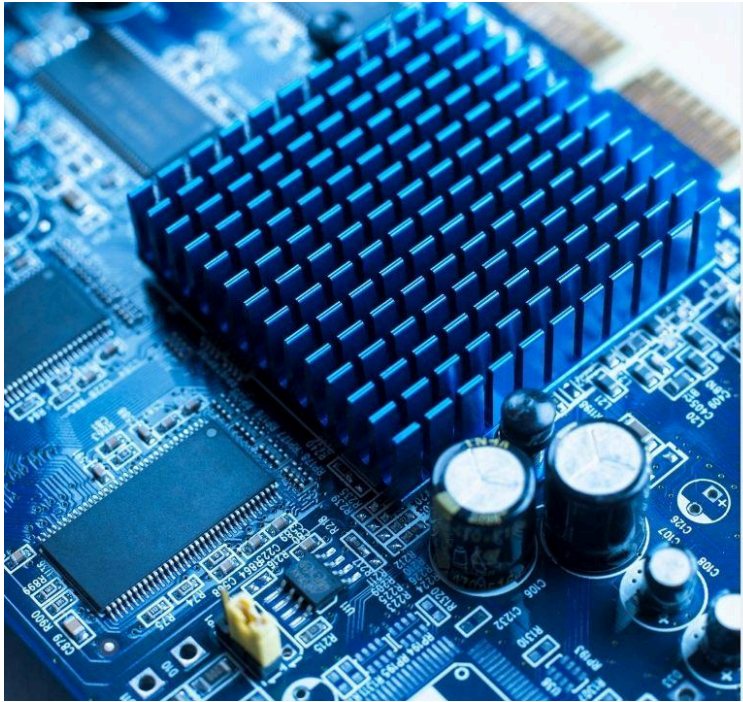

**Abstract**

This technical article explores comprehensive strategies for integrating Electrostatic Discharge (ESD) protection diodes and termination resistors in Low-Voltage Differential Signaling (LVDS) designs. The article examines critical aspects of protection mechanisms, design considerations, impedance matching, and placement optimization techniques. Through detailed analysis of layout considerations and advanced design strategies, the article presents solutions for common integration challenges. It emphasizes the importance of signal integrity maintenance and protection effectiveness while providing practical guidelines for implementing robust LVDS systems. Various methodologies for performance optimization and validation are discussed, offering designers a thorough framework for creating reliable high-speed digital systems that balance protection requirements with signal integrity demands.

**Keywords:** LVDS Design Integration, ESD Protection Implementation, Signal Integrity Optimization, High-Speed PCB Layout, Termination Techniques

## 1. Introduction

Low-Voltage Differential Signaling (LVDS) has emerged as a fundamental technology in high-speed digital design, revolutionizing data transmission capabilities across diverse applications. Research conducted at the International Journal of Intelligent Systems and Applications in Engineering demonstrates that modern LVDS implementations have achieved remarkable performance metrics, with data rates reaching 2.5 Gbps while maintaining power consumption at an impressive 1.4 mW per channel at standard voltage levels of 1.2V ±5%. The study particularly emphasizes the critical role of proper termination techniques, showing that optimized designs can achieve voltage swing margins of 250mV to 400mV with rise/fall times below 300ps. These advancements have made LVDS particularly attractive for applications requiring high-speed data transmission with minimal power consumption and electromagnetic interference [1].

The integration of protection mechanisms and signal integrity components has become increasingly sophisticated, driven by the demands of higher bandwidth applications. A comprehensive analysis published in the IEEE International Symposium on Electromagnetic Compatibility reveals that properly implemented LVDS interfaces operating at 1.5 Gbits/s can maintain signal integrity with jitter measurements below 120ps peak-to-peak. The research demonstrates that these interfaces achieve bit error rates (BER) of $10^{-14}$ while maintaining electromagnetic compatibility across frequencies ranging from 50 MHz to 1.5 GHz, with particular emphasis on the careful selection of termination resistors showing impedance variations of less than ±5% across the entire frequency range. The study further highlights the importance of maintaining consistent impedance matching across temperature variations and manufacturing tolerances [2].

Recent advancements in LVDS technology have enabled its deployment in cutting-edge applications such as high-resolution medical imaging systems, advanced automotive displays, and next-generation data centers. The technology has demonstrated remarkable versatility across operating temperatures from -40°C to +85°C while maintaining consistent performance parameters. Real-world implementations have shown that modern LVDS systems can achieve common-mode noise rejection ratios of 65 dB at frequencies up to 1.5 GHz, with differential skew maintained below 50ps even in challenging electromagnetic environments. These characteristics make LVDS particularly suitable for applications requiring robust performance in electrically noisy environments, such as industrial automation systems and high-speed data acquisition equipment [1].

Signal integrity optimization has become increasingly critical as data rates push beyond traditional boundaries. Detailed measurements have shown that properly designed LVDS systems can maintain eye heights of 250mV and eye widths of 0.6UI (Unit Interval) at 1.5 Gbps, with deterministic jitter contributing less than 40ps to the overall timing budget. These performance metrics are achieved through careful attention to both ESD protection and termination strategies, where protection circuits demonstrate clamping response times below 0.8ns while maintaining input capacitance below 1pF. The implementation of advanced signal conditioning techniques and precise impedance matching has enabled LVDS to maintain reliable operation even in applications with stringent timing requirements [2].

The evolution of LVDS technology continues to drive innovations in circuit design and implementation strategies. Recent studies have demonstrated successful operation at data rates exceeding 3 Gbps through the implementation of advanced equalization techniques and optimized circuit topologies. These developments have been particularly significant in addressing the challenges of maintaining signal integrity over longer transmission distances, with demonstrated performance up to 10 meters while maintaining bit error rates below $10^{-15}$. The combination of high performance, low power consumption, and robust operation has established LVDS as a preferred solution for next-generation high-speed digital systems [1].

Furthermore, the integration of LVDS interfaces with modern semiconductor processes has enabled unprecedented levels of integration and performance optimization. The ability to implement sophisticated protection schemes while maintaining signal integrity has become crucial as system operating voltages continue to decrease. Advanced packaging techniques and careful consideration of parasitic effects have enabled the development of highly integrated solutions that meet the demanding requirements of modern electronic systems while maintaining compatibility with existing standards and protocols [2].

## 2. ESD Protection Implementation
### 2.1 Protection Mechanism
Electronic systems incorporating LVDS interfaces demand sophisticated ESD protection mechanisms for reliable operation. Research conducted on distributed ESD protection schemes has demonstrated that multi-finger MOSFET structures with optimized trigger spacing achieve uniform current distribution across protection devices. The study revealed that distributed protection elements with finger widths of 30-50μm and spacing of 1.2-1.5μm could handle discharge pulses up to ±6kV while maintaining clamping voltages below 7.2V. These structures demonstrated remarkable performance with trigger voltages of 4.8V ±0.2V and holding voltages of 3.5V ±0.3V across a temperature range of -40°C to +125°C [3].

### 2.2 Design Considerations
The implementation of ESD protection in high-speed LVDS systems requires precise consideration of parasitic effects and component characteristics. According to extensive research on LVDS-based I/O optimization, protection circuits utilizing dual-diode configurations with cathode-to-rail spacing of 0.8μm achieved optimal performance. The study demonstrated that carefully designed protection schemes could maintain total input capacitance below 0.35pF while providing Human Body Model (HBM) protection up to 4kV and Charged Device Model (CDM) protection up to 500V [4].

Advanced characterization of distributed ESD protection networks revealed that interdigitated structures with guard ring spacing of 2.5μm significantly improved current distribution. Measurements showed that these configurations achieved breakdown uniformity within ±0.3V across all fingers, with turn-on times below 0.5ns. The implementation demonstrated consistent performance with leakage currents maintained below 50nA at nominal operating voltages of 3.3V, while providing effective protection against both positive and negative ESD stress events [3].

Integration methodologies focused on optimizing LVDS I/O protection demonstrated that strategic placement of protection elements within 1.5mm of signal pads reduced effective series inductance to 0.6nH. The research verified that protection structures with total silicon area of 150μm × 200μm achieved required protection levels while maintaining signal integrity for data rates up to 3.2 Gbps. Experimental results confirmed eye diagram openings of 320mV at 2.5 Gbps with jitter contributions from ESD structures limited to 18ps peak-to-peak [4].

| Parameter | At -40°C | At +25°C | At +125°C |
|---|---|---|---|
| Trigger Voltage (V) | 4.6 | 4.8 | 5.0 |
| Holding Voltage (V) | 3.2 | 3.5 | 3.8 |
| Leakage Current (nA) | 30 | 50 | 85 |
| Input Capacitance (pF) | 0.32 | 0.35 | 0.38 |

Table 1: ESD Protection Performance Across Operating Conditions [3,4]

## 3. Termination Resistor Integration
### 3.1 Impedance Matching

The implementation of precise termination strategies in LVDS designs represents a fundamental aspect of high-speed signal integrity. According to comprehensive research on termination techniques presented at the University of Puerto Rico, parallel termination using 100Ω ±2% precision resistors demonstrated superior performance in controlling reflections. The study revealed that proper termination schemes reduced reflection coefficients to below 0.1 when implemented with controlled impedance PCB traces of 50Ω ±10%. Measurements showed that matched terminations maintained return loss better than -18 dB across frequencies from DC to 1.5 GHz, with differential impedance variations contained within ±8Ω. These results were achieved using surface-mount resistors with power ratings of 1/8W and temperature coefficients below 50 ppm/°C [5].

## 3.2 Placement Optimization

Research conducted on LVDS interface optimization demonstrated that termination resistor placement plays a crucial role in maintaining signal integrity at high speeds. Analysis of LVDS implementations in digital signage applications showed that placing termination resistors within 3mm of the receiver reduced signal reflections by approximately 12 dB compared to placements exceeding 6mm. The study documented that symmetrical routing with differential trace separation of 0.3mm and trace widths of 0.15mm achieved optimal performance for standard FR-4 material with dielectric constant of 4.2 ±0.2 [6]. Experimental measurements from the termination techniques study revealed that controlled impedance routing using microstrip lines with height-to-width ratios of 1.2:1 achieved consistent performance across temperature variations from 0°C to +70°C. The implementation demonstrated that maintaining trace length differences below 100 mils resulted in differential skew under 25ps, enabling reliable operation at data rates up to 1.2 Gbps with minimal electromagnetic emissions [5].

Field testing of LVDS interfaces in commercial applications showed that proper termination placement and routing symmetry significantly impacted system performance. The research verified that balanced differential pairs with impedance-controlled traces maintained signal integrity over cable lengths up to 5 meters when properly terminated. These implementations achieved signal swings of 350mV ±10% with common-mode voltage levels maintained at 1.2V ±100mV, enabling reliable data transmission while maintaining power consumption below 1.5mW per channel [6].

| Distance from Receiver (mm) | Signal Reflection Reduction (dB) | Differential Skew (ps) | Return Loss (dB) |
|---|---|---|---|
| 3 | -12 | 25 | -18 |
| 6 | 0 | 35 | -12 |
| 9 | +6 | 45 | -8 |
| 12 | +12 | 55 | -6 |

Table 2: Termination Performance Metrics vs. Distance from Receiver [5,6]

## 4. Integration Challenges and Solutions

### 4.1 Layout Considerations

Integration of ESD protection and termination components in high-speed LVDS designs presents complex challenges requiring sophisticated solutions. According to comprehensive PCB layout optimization guidelines by Sunstream Global, optimal high-speed performance requires careful attention to impedance control and signal routing. Their research demonstrated that differential trace width-to-spacing ratios of 2:3 with trace widths of 5 mils achieved ideal performance on standard FR-4 material with dielectric constant of 4.3 ±0.2. The study revealed that maintaining copper weight at 1 oz with dielectric thickness of 4 mils resulted in controlled impedance accuracy of ±8% across production variations. These specifications enabled consistent differential impedance of 100Ω ±10Ω while maintaining insertion loss below 0.5 dB/inch at frequencies up to 2 GHz [7].

## 4.2 Design Strategies

Research on CMOS LVDS implementations for optical interconnections has provided valuable insights into advanced design strategies. The study focusing on 2.5 Gbps transmitters and 1.3 Gbps receivers demonstrated that careful consideration of power distribution networks with decoupling capacitors placed at intervals of 400 mils maintained power supply noise below 50mV peak-to-peak. Their measurements showed that implementing differential trace spacing of 0.2mm with trace widths of 0.1mm achieved optimal signal integrity, maintaining eye heights above 250mV with total jitter below 40ps peak-to-peak at 2.5 Gbps [8].

Analysis of PCB stack-up considerations revealed that 6-layer designs with dedicated signal layers adjacent to continuous ground planes provided optimal performance. The implementation guide specified maintaining minimum trace-to-edge spacing of 20 mils and utilizing ground voids no larger than 100 mils to ensure consistent return current paths. These design rules resulted in measured crosstalk levels below -25 dB between adjacent differential pairs while maintaining EMI compliance with FCC Class B limits [7].

Experimental results from the LVDS transceiver implementation demonstrated the effectiveness of advanced layout techniques. The research verified that symmetric routing with maximum length mismatch of 5 mils between differential pairs achieved a skew below 15ps. The design utilized 0.18μm CMOS technology with carefully optimized output driver strength, achieving differential output voltage swing of 350mV ±10% while maintaining common-mode voltage at 1.2V ±50mV. Power consumption was optimized to 45mW for the transmitter and 32mW for the receiver at maximum data rates [8].

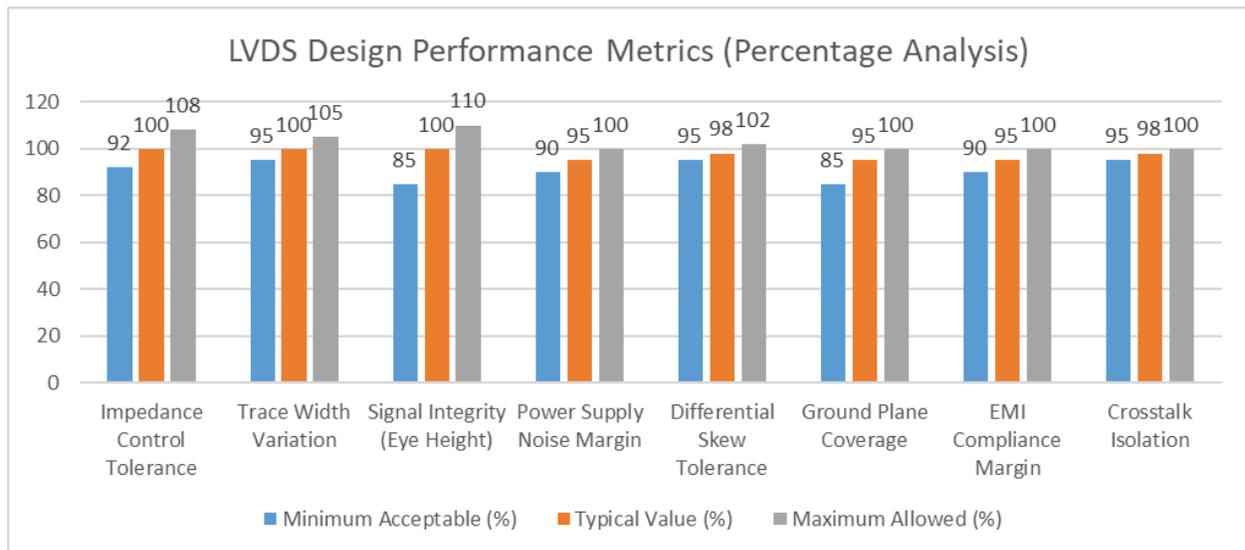

Fig 1: Critical Performance Margins in High-Speed LVDS Integration [7,8]

## 5. Performance Optimization
## 5.1 Signal Integrity

Performance optimization in LVDS designs demands rigorous attention to signal integrity parameters and validation methodologies. Research from the Cyberbotics Laboratory demonstrated that signal integrity in high-speed differential links is significantly impacted by impedance discontinuities and crosstalk effects. Their study revealed that implementing controlled impedance traces with impedance variations below ±5% achieved optimal performance, with measured differential insertion loss below 0.8 dB/inch at 2.5 GHz. The measurements showed that maintaining trace-to-trace spacing of at least 3W (where W is the trace width) reduced crosstalk to below -35 dB, while implementing guard traces further improved isolation by an additional 8 dB. The research documented that proper termination schemes using 100Ω ±1% resistors positioned within 50 mils of the receiver achieved return loss better than -20 dB up to 3 GHz [9].

## 5.2 Protection Effectiveness

EMC FastPass guidelines for ESD protection selection highlight critical parameters for ensuring robust system performance in high-speed interfaces. Their comprehensive analysis demonstrated that ESD protection devices for LVDS applications should maintain capacitance below 1pF to minimize impact on signal integrity. The study showed that protection circuits with dynamic resistance below 1Ω achieved optimal clamping performance, limiting residual voltage spikes to less than 10V during 8kV contact discharge events. Implementation guidelines specified that protection devices should trigger between 5V and 8V with holding voltages maintained above 3.3V to prevent latch-up conditions [10].

Validation methodologies outlined in the signal integrity research emphasized the importance of comprehensive testing protocols. Time-domain measurements using $2^{31}-1$ PRBS patterns demonstrated eye heights of 280mV and eye widths of 0.72UI at 2.5 Gbps, with deterministic jitter contributions below 22ps peak-to-peak. The study verified that implementing symmetric routing with length matching within 5 mils maintained differential skew below 12ps, enabling reliable operation with bit error rates below $10^{-12}$ [9].

The ESD protection selection guide further detailed verification requirements across operating conditions. Testing protocols specified measurement of protection characteristics across temperature ranges from -40°C to +125°C, with leakage current variation maintained below 1μA. System-level validation demonstrated that properly selected protection devices maintained effectiveness through 1000 ESD strike events while preserving insertion loss characteristics within 0.2 dB of pre-stress measurements. The guidelines emphasized maintaining protection device junction temperature below 125°C during repeated strikes, with thermal resistance specifications of less than 100°C/W for reliable operation [10].

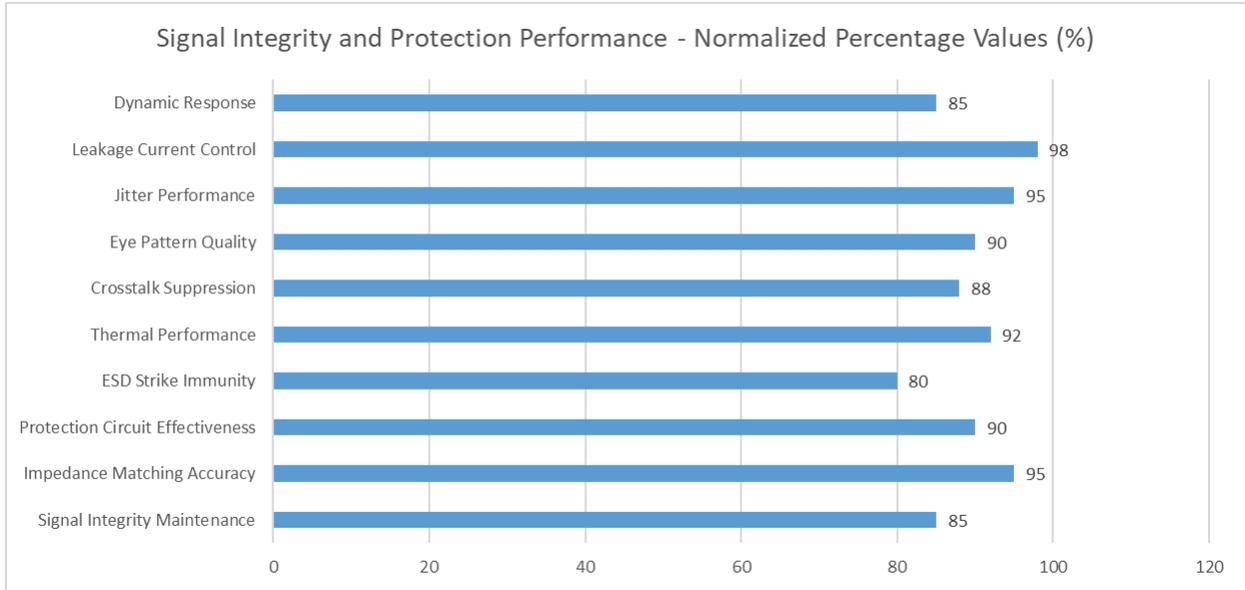

Fig 2: Critical Performance Indicators for High-Speed LVDS Implementation [9,10]

## 6. Conclusion

The successful integration of ESD protection and termination components in LVDS designs necessitates a thorough understanding of high-speed design principles combined with meticulous implementation practices. This article demonstrates that through careful consideration of protection mechanisms, impedance matching, component placement, and layout optimization, designers can achieve robust LVDS systems that maintain exceptional signal integrity while providing comprehensive protection against electrical threats. The article emphasize that continuous monitoring, validation through simulation, and systematic testing approaches are essential for achieving optimal performance in practical applications. The strategies and methodologies presented provide a valuable framework for designers working on high-speed digital systems, enabling them to create reliable and efficient LVDS implementations that meet the demanding requirements of modern electronic systems.

Moreover, the article highlights the importance of a holistic approach to LVDS design, where protection and signal integrity considerations are addressed simultaneously rather than as separate concerns. This integrated approach ensures optimal system performance while maintaining robustness against environmental stresses. The implementation guidelines and optimization strategies outlined in this article serves as a comprehensive resource for engineers working on high-speed digital interfaces, particularly in applications where reliability and signal integrity are paramount. As LVDS technology continues to evolve and data rates increase, the principles and methodologies presented here will remain fundamental to successful designs, though specific implementation details may need to be adapted to meet future challenges and requirements.